# Prostate cancer histopathology with label-free multispectral deep UV microscopy quantifies phenotypes of tumor grade and aggressiveness

Soheil Soltani[1], Ashkan Ojaghi[1], Hui Qiao[2], Nischita Kaza[3], Xinyang Li[2], Qionghai Dai[2], Adeboye O Osunkoya[4, 5], Francisco E Robles[1,3,5]*

**Affiliations:**

[1]Wallace H. Coulter Dept. of Biomedical Engineering, Georgia Institute of Technology and Emory University, Atlanta, GA 30332, USA

[2]Department of Automation, Tsinghua University, Beijing 100084, China

[3]School of Electrical and Computer Engineering, Georgia Institute of Technology, Atlanta, Georgia 30313, USA

[4]Departments of Pathology and Urology, Emory University School of Medicine, Atlanta, GA 30322

[5]Winship Cancer Institute of Emory University, Atlanta, GA 30322

*Corresponding author Email: robles@gatech.edu

**Quantitative Grading of Prostate Cancer:** Leveraging multispectral deep-UV microscopy, a label-free molecular imaging technology, we introduce a novel approach that enables quantitative grading of thin prostate cancer tissue sections. Using endogenous light scattering and absorption UV spectral signatures, we define several "optical stains" which highlight differences among tissue structures, including nuclei, cytoplasm, stroma, basal cells, nerves, inflammation, and most importantly, malignant glandular phenotypes, with subcellular spatial resolution. Lastly, using state-of-the-art deep learning methods, we translate our label-free, quantitative deep-UV images into virtual H&E images with near perfect agreement to the gold standard H&E-stained tissue sections. This work has significant implications for improving prostate cancer diagnosis and management.

**Abstract:** Identifying prostate cancer patients that are harboring aggressive forms of prostate cancer remains a significant clinical challenge. To shed light on this problem, we develop an approach based on multispectral deep-ultraviolet (UV) microscopy that provides novel quantitative insight into the aggressiveness and grade of this disease. First, we find that UV spectral signatures from endogenous molecules give rise to a phenotypical continuum that differentiates critical structures of thin tissue sections with subcellular spatial resolution, including nuclei, cytoplasm, stroma, basal cells, nerves, and inflammation. Further, we show that this phenotypical continuum can be applied as a surrogate biomarker of prostate cancer malignancy, where patients with the most aggressive tumors show a ubiquitous glandular phenotypical shift. Lastly, we adapt a two-part Cycle-consistent Generative Adversarial Network to translate the label-free deep-UV images into virtual hematoxylin and eosin (H&E) stained images. Agreement between the virtual H&E images and the gold standard H&E-stained tissue sections is evaluated by a panel of pathologists who find that the two modalities are in excellent agreement. This work has significant implications towards improving our ability to objectively quantify prostate cancer grade and aggressiveness, thus improving the management and clinical outcomes of prostate cancer patients. This same approach can also be applied broadly in other tumor types to achieve low-cost, stain-free, quantitative histopathological analysis.

**Introduction**

Prostate cancer (PCa) is the most commonly diagnosed (non-cutaneous) cancer among men in the United States (1). According to the National Institute of Health SEER report, over 3.25 million men in the US are currently living with this disease, and ~1 in 6 men will be diagnosed with it over their lifetime (2). Other estimates, however, suggest that the prevalence of PCa may actually be much higher. Studies using autopsy analyses indicate that over half of all men above the age of 50 harbor some form of PCa, increasing the estimate of the number of men living with this disease to over 20 million in the US (3). This staggering prevalence makes it clear that in a significant number of cases PCa follow an indolent course; nevertheless, PCa is still the second leading cause of cancer death in men, with over 33,000 deaths in the US in 2020 (4). Thus, while early and accurate PCa detection is critical, so too is the ability to objectively assess the tumors' aggressiveness. Unfortunately, this remains a significant clinical challenge which has profound implications. On the one hand, there are a vast number of PCa patients harboring indolent tumors who are either (i) under a 'watchful waiting' category (i.e., wait and see if the cancer progresses) and have to live with the uncertainty of potentially having an unsampled aggressive tumor, or (ii) deemed high risk but are ultimately over-diagnosed and over-treated. On the other hand, there are many PCa patients harboring aggressive tumors who are at risk of being under-diagnosed and under-treated—for these patients an incorrect diagnosis may be fatal.

In the current standard of care, once a diagnosis of PCa has been established, one of the most important factors in assessing tumor aggressiveness is the Gleason score (this also largely dictates treatment course) (5-7). Here a histopathologist visually inspects hematoxylin and eosin (H&E) stained thin tissue sections and determines the two most common glandular/architectural patterns, which are assigned a grade from 3 to 5 (grades of 1 and 2 are not diagnosed on needle core biopsies, and are non-cancerous). While accepted as the gold standard, the Gleason score (the sum of the two grades) is qualitative and subject to intra- and inter-observer variability. Some studies have reported inter-observer agreement (kappa values) ranging from ~40% to 70%, with a significant portion of the discordant values dictating different treatment paths (8-10). H&E staining can also be highly variable and laboratory dependent, which undoubtedly contributes to observer variability, but also makes it challenging to extract quantitative parameters. Thus, there is a significant need for novel technologies that can provide pathologists and clinicians with additional quantitative information regarding the aggressiveness of PCa, and thus prognosis for individual patients.

Advanced method using genetic profiling, for example, provide a wealth of information but have shown limited success in predicting the prognosis of cancer patients (11, 12). The poor predictive power may be attributed to the vast genetic heterogeneity of tumors, which makes it extremely difficult to identify a unique set of mutations that provide reliable prognostic information. Alternatively, recent efforts have shifted towards exploring phenotypical "common-denominators" to the countless genetic and epigenetic alterations that lead to cancer (13-29). Phenotypical changes, including changes in metabolites, nuclear morphology, and nano-architecture, are more consistent across patients than the myriad of individual mutations and disrupted pathways underlying the disease, and can potentially better characterize tumors. This approach has shown very promising results for early cancer detection (13-26), and—to a more limited extent—assessing cancer aggressiveness (27-29).

Here we apply multispectral deep-UV microscopy as a novel tool for phenotyping PCa tissue sections, resulting in a unique quantitative biomarker that correlates with Gleason score (Grade group) and that is predictive of disease aggressiveness. Multispectral deep-UV microscopy offers rich endogenous, label-free, molecular information of important tissue biomolecules with subcellular spatial resolution using a fast, low-cost imaging configuration (30-35). To define this novel biomarker (or phenotypical continuum), we couple deep-UV microscopy with an unsupervised analysis of the molecular signatures (28). Importantly, we find that patients with the most aggressive forms of prostate cancer express a ubiquitous glandular phenotypical shift, even in glands that appear to be less aggressive. We further introduce multiple virtual "optical stains" (or "biochemical stains") of tissue slides that highlight important components for disease diagnosis such as nuclei, cytoplasm, stroma, basal layer, nerves, and inflammation. The unique insight provided by the method is not available with current histological methods. Finally, we leverage recent advances in deep learning to translate our multi-spectral deep UV images into virtual H&E-stained images which show a high degree of correlation with the gold-standard H&E histopathological images of prostate tissue. Results from this work have significant implications towards improving diagnosis and management of prostate cancer. Further, this same approach may be widely applicable to improve histopathological analysis in many other tissue types and diseases.

**Results**

**Deep-UV microscopy of prostate tissue sections**

Details of the multispectral deep-UV microscope are provided in the methods and materials section. Images were acquired from unlabeled fixed radical prostatectomy tissue samples, which were sliced (~5μm thick) and mounted on quartz microscope slides. Images were acquired from histologically important regions containing structures with benign tissue, inflammation, stroma, high grade prostatic intraepithelial neoplasia (HGPIN), and glands with various grades of prostate cancer (Gleason grades 3, 4, and 5). Eighty-seven regions of interest were acquired from 15 patients. Each region was ~1mm X 1.5mm, acquired with a spatial resolution of ~250nm. Multispectral images were taken at four key wavelengths, including 220nm, 255nm, and 280nm, which correspond to strong absorption peaks of proteins and nucleic acids (35). We also included 300nm which does not correspond to an absorption peak of any endogenous biomolecule but serves as an indicator of tissue scattering, which has been applied as a surrogate biomarker of tissue nano-architecture (24, 26, 35-42).

The multispectral data were processed using a geometrical representation of principal component analysis (PCA), an unsupervised method (28). In this process, approximately 130 million spectra from select representative regions were used to calculate the principal components (PCs). Figure 1a shows the resulting orthogonal PCs. It is important to note that these vectors, while purely mathematical in nature, in fact resemble the absorption and scattering spectral behavior of biological tissues (35). For example, the first principal component, PC1, shows a unipolar, monotonically decreasing behavior that is consistent with the expected response of tissue scattering. PC2 and PC4 show peak responses that correspond to protein absorption, while PC3 shows an inverted peak that is in agreement with the absorption from nucleic acid (35). Nevertheless, projections of the spectra onto these PCs do not uniquely correspond to these molecules, and do not prominently highlight important tissue structures alone, as seen in Fig 1c.

To obtain a more natural representation of the endogenous tissue composition, we transform these data from a Cartesian coordinate system with only the first three PCs (which possess over 99% of the data variance) to spherical coordinates (see Fig 1d). (The same procedure can be applied with any combination of three PCs.) In this representation, the azimuth ($\theta$) and elevation ($\phi$) angles contain all the information about the shape of the spectra; in other words, these two dimensions contain nearly all the available biophysical and biochemical information. The radius, on other hand, serves as a relative measure of the concentration. Thus, images can be represented in a hue-saturation-value (HSV) color space, with the hue given by the angular coordinates (either elevation or azimuth angle, as shown in Fig. 1e), the value set by the radius, and the saturation fixed to 1.

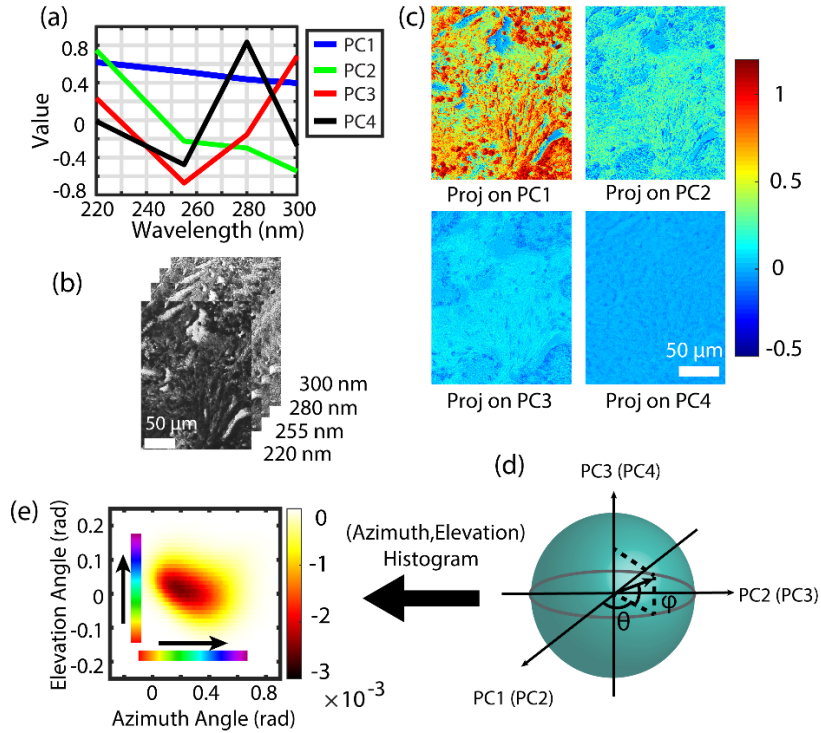

Figure 1. Processing of multi-spectral deep UV images using a geometrical representation of PCA. (a) The 4 principal components resulting from 130 million spectra from representative select regions. (b) A representative multi-spectral deep UV transmission data cube taken at 220, 255, 280 and 300 nm. (c) Calculated projections of the data cube on the principal components. (d) Schematic of data conversion from Cartesian to Spherical coordinates. (e) Representative 2D histogram of a data cube using azimuthal and elevation coordinates.

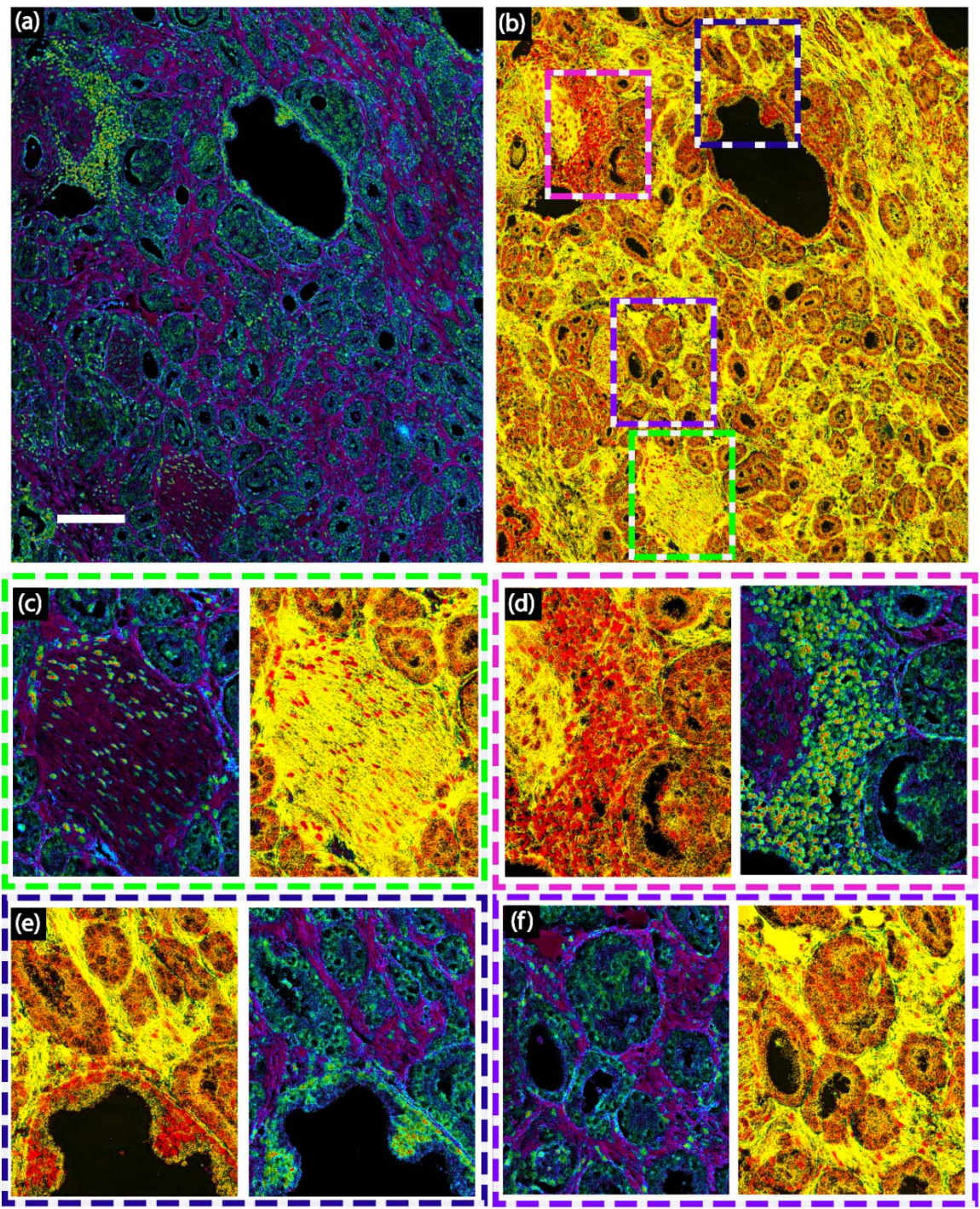

Figure 2. A representative demonstration of two "optical stains". Colorization scheme using (a) principal components 1,2 and 3 in Elevation direction and (b) principal components 2,3 and 4 in Azimuth direction. Scale bar: 150 μm. Comparison of the two color-coding schemes for (c) a nerve, (d) a region with inflammation (e) an entrapped benign prostate gland surrounded by Gleason Grade 3 and 4 cancer glands. (f) Prostate cancer glands with Gleason Grade 3 next to a Gleason Grade 4 glomeruloid gland. All rectangles are 210μm x260μm

An important feature of this geometrical PCA representation is that each point in the 2D histogram of the elevation and azimuth angles (Fig. 1e) represents a unique spectral response, and hence different molecular and/or biophysical makeup. Thus, by color-coding images based on the angular distributions, we are able to assign a unique hue to spatial regions with similar composition. The resulting "optical stains" enhance contrast among various structures in prostate tissue sections which can be leveraged, along with H&E, for diagnostic applications.

Figure 2 shows two types of "optical stains" that highlight important tissue structures. In the first (Fig. 2a), the elevation angle is used to encode hue which yields the most prominent contrast for cell nuclei, depicted in green. This is consistent with the general behavior of the PCs, as the elevation angle in this case corresponds to a ratio of the 3rd PC—which resembles the inverted absorption peak from nucleic acids—relative to both the 1st and 2nd PCs (which correlate with scattering and protein spectral signatures, respectively). Thus, nuclei are mapped to regions with negative elevation angles. Further, in this representation, the stroma shows a dark purple color (and has a positive elevation angle). In the second colorization scheme (Fig. 2b), we encode hue based on the azimuthal angle derived from a 3D space from the 2nd, 3rd and 4th principal components. Here the hue encodes differences between proteins and nucleic acid, without contributions from scattering (1st PC). The resulting images (Fig 2b) exhibit some degree of nuclear contrast (depicted in red), but most prominently show the stroma in bright yellow. Figures 2c-f highlight selected regions from a nerve surrounded by prostate cancer glands (Fig. 2c), a highly inflamed region (Fig. 2d), an entrapped benign prostate gland next to cancer glands (Fig. 2e), and prostate cancer glands (Fig. 2f).

Figure 3 shows additional examples that emphasize the ability of these label-free optical stains to provide unique contrast among different structures, including benign tissues, HGPIN, Gleason grades3-5, necrosis, inflammation, and even red blood cells. Images from H&E-stained tissues (from adjacent sections) are also shown for comparison. While not in perfect one-to-one agreement, in general, the overall tissue structure observed with H&E is preserved in the label-free UV images, including clear contrast between nuclei and stroma. An important distinction, however, is that the information derived from UV microscopy is quantitative. Further, with the UV images, subtle differences in hue can be observed in the various structures, including glands with different Gleason grades, inflammation, necrosis, and HGPIN.

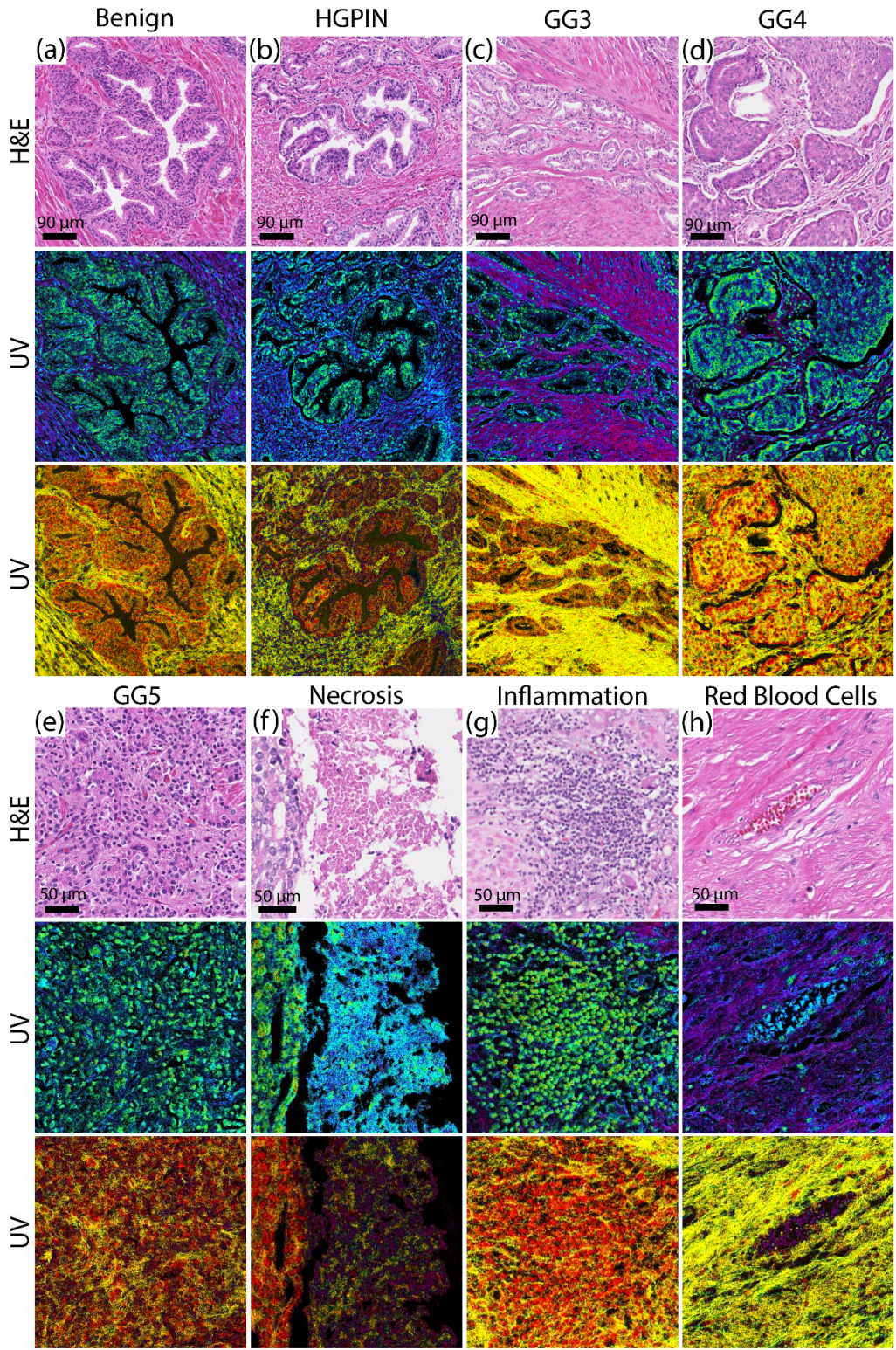

Figure 3: Comparison of the two "optical stains" with corresponding H&E-stained tissue scans from various prostate tissue structures. (a) Benign gland. (b) High grade Prostatic intraepithelial neoplasia (PIN) region. (c) Cancer region with Gleason Grade 3 glands. (d) Cancer region with Cribriform Gleason Grade 4 region. (e) Region with Gleason Grade 5. (f) Region with necrosis inside a Gleason Grade 5

cancer gland. Necrosis is clearly distinguishable from the cancer cells on the left side of the image. (g) Inflammation. (h) Red blood cells.

**Prostate cancer diagnosis and grading using deep-UV microscopy**

Using a 3D space defined by the first three PCs, we defined a third "optical stain" by encoding hue using the azimuth angle (Fig. 4). These maps highlight contributions from light scattering (from PC1) relative to both proteins and nucleic acid (PC2 and PC3, respectively). We note that scattering variations arise from genetic and epigenetic perturbations that results in micro and/or nano-scale alterations in intracellular milieu, such as the cytoskeleton, ribosomes, chromatin, mitochondria, and collagen fibrils that are known to be altered in field carcinogenesis (36-41). Furthermore, protein and nucleic acid alterations have also been well documented throughout the progression of prostate cancer (43-51).

The resulting image representation (i.e. optical stain) does not exhibit contrast to structures conventionally used in histopathology (e.g. nuclei, cytoplasm, stroma, etc.); instead, we find that this representation encodes for a glandular phenotype that correlated with malignancy. Figure 4 shows two examples from patients with intermediate-grade cancer. Here benign glands possess a blue hue, while glands with cancer (Gleason grades 3 or 4) exhibit a relative shift captured in green to red hues which represents an increase in nucleic acid and protein content, potentially from cell overgrowth byproducts (43-46, 49, 52-54). In these maps, the glands were manually segmented for clarity. Again, the change in color represents alterations in the scattering properties relative to protein and nucleic acid content, all of which have been implicated in early-stage alterations of cancer, as well as metastatic disease (36, 37, 39, 41, 48) . Thus, the azimuth angle from a geometrical representation of the first three PCs effectively yields a phenotypical continuum that can be applied as a surrogate biomarker of prostate cancer malignancy.

It is worth highlighting important features in Fig. 4. Figures 4a-b show a set of pseudo-neoplastic benign glands (blue arrows) that are not well formed, meaning they express slight cytological and morphological variations such as cytoplasm clearing that classifies them as a mimicker of prostatic adenocarcinoma (typically of Gleason grade 3). However, the existence of basal cells around the glands as well as the papillary infoldings of the gland differentiates them from carcinoma. And indeed, the malignancy optical stain clearly indicates that these glands are benign and distinct from Gleason grade 3 and 4 glands (green and red arrows, respectively). Figures 4c-d show benign central zone histology glands (blue arrows) surrounded by Gleason Grade 3 cancer glands (green arrows). Central zone histology glands are potential mimickers of HGPIN and Gleason Grade 4 cancer glands (Cribriform) and are often difficult to differentiate from cancer glands (55, 56); nevertheless, the malignancy optical stain identifies these glands as benign. Further, Fig. 4d clearly shows a gradual color gradient, and hence phenotypical continuum, from left to right as the glands progress from benign to cancer. It is clear that the information provided by this optical stain is independent and complimentary to the gold standard H&E stain.

Supplemental Fig. S2 shows additional examples from patients with aggressive disease (i.e., those containing Gleason grade 5 glands). These samples possess a unique response which is discussed in more detail below.

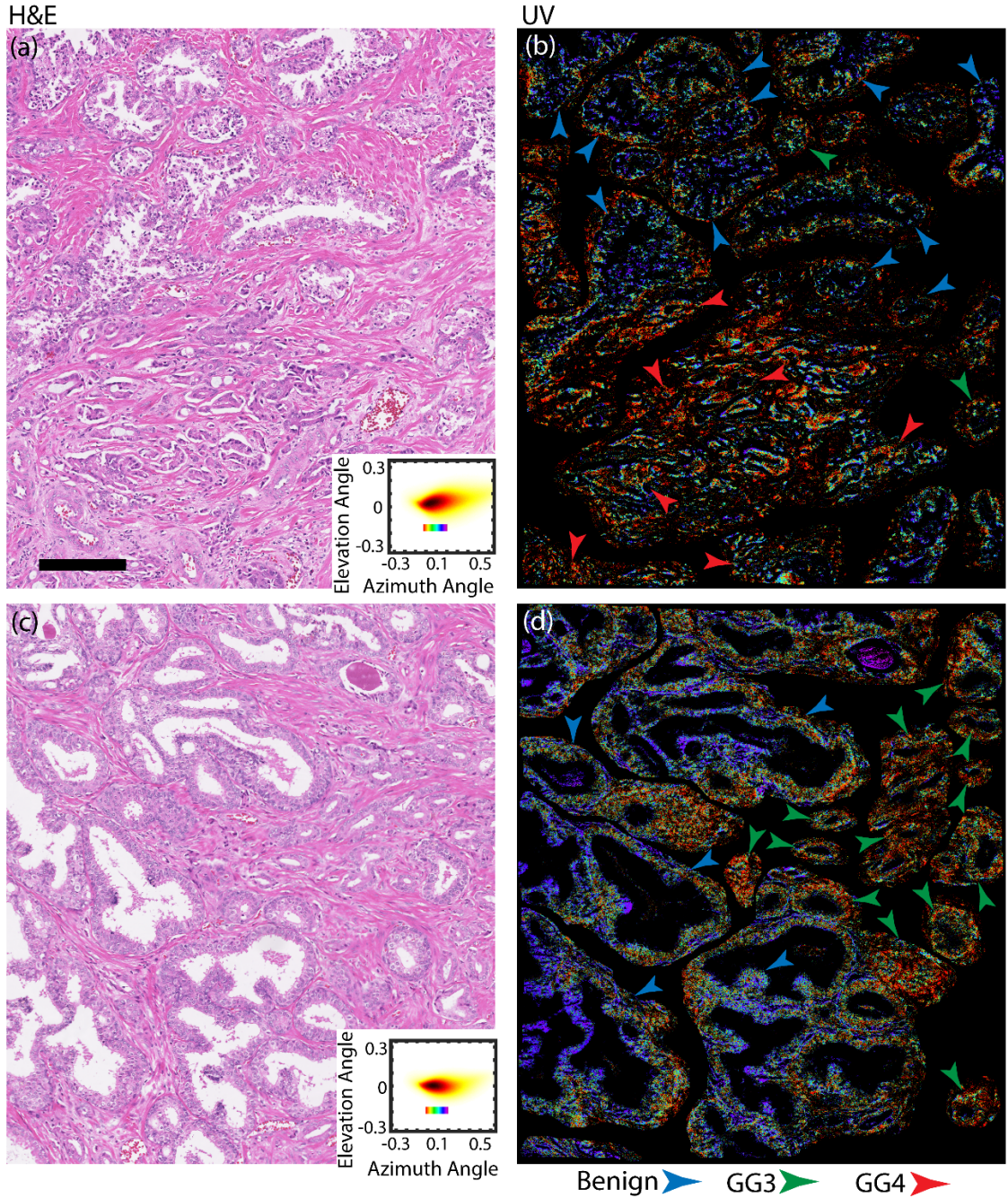

Figure 4: Comparison of malignancy maps from different prostate regions with corresponding H&E scans. Insets show the 2D histograms for comparison. As clearly evident from Fig. 4 the malignancy optical stain shows diagnostic capabilities complimentary to H&E, where only morphological parameters are considered. In Fig. 4 (b) and (d) we have manually removed stroma and inflammation regions to aid visibility. (Scale bar: 200 μm)

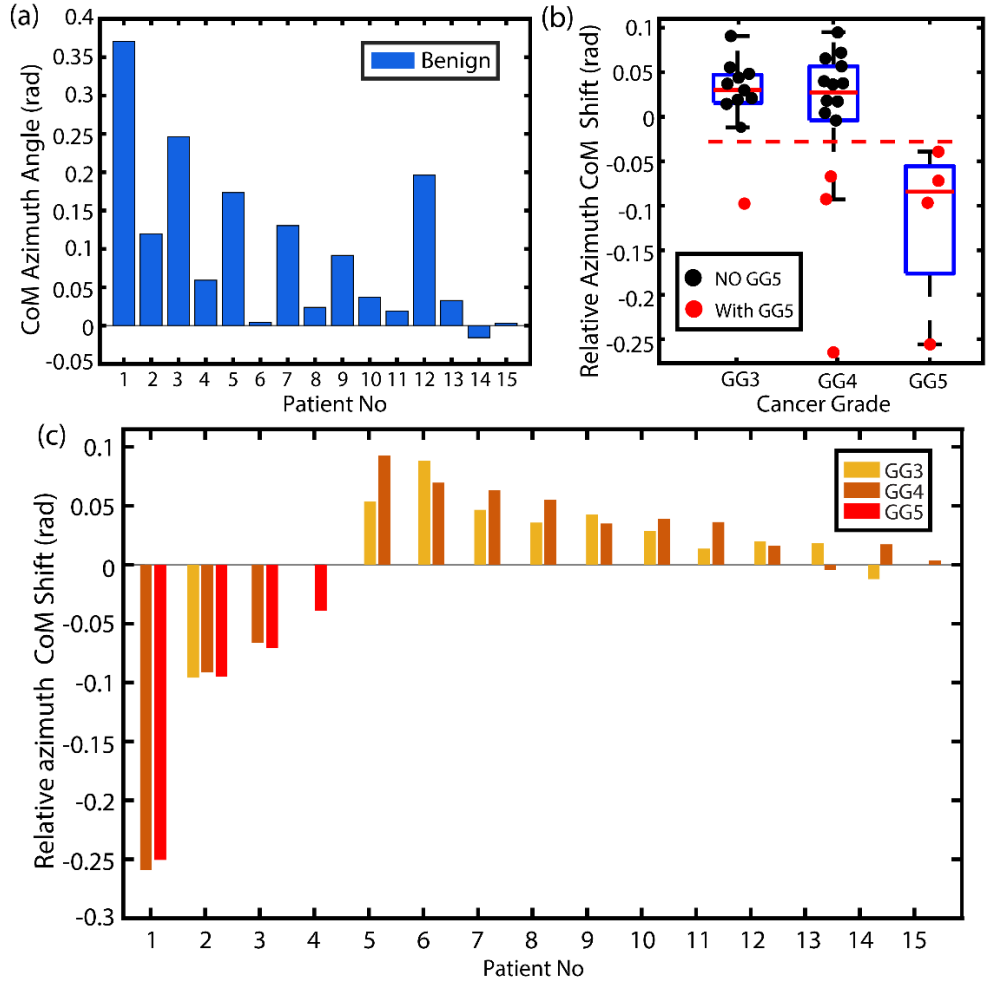

Figure 5. Absolute and relative CoM azimuthal angles serves as a personalized malignancy biomarker and reveal unique glandular phenotypes. (a) Comparison of absolute maximum peak azimuthal coordinates of integrated histograms of benign regions for 15 patients. (b) Cumulative boxplots for calculated relative azimuthal shift for different prostate cancer grades in 15 patients. (c) Barplots for calculated relative azimuthal shift for different prostate cancer grades in 15 patients with the benign region used as reference for each patient.   In Figs. 5(b) and (c) it is clear that the more aggressive phenotypes have an opposite shift even for lower grades of cancer. In Fig. 5 (b) the red dotted line is the threshold of the shift for the more aggressive cancer regions on the opposite direction.

To investigate the properties of this phenotypical shift further, we analyze the cumulative behavior of benign glands and cancerous glands with the same Gleason grade for each patient. In this process, cumulative 2D histograms were generated for each type of gland (benign and Gleason grades 3-5) for each patient, then data were integrated across elevation angle, and finally the center of mass (CoM) of the resulting azimuth angle distributions were computed. This value effectively quantifies the hues in the malignancy optical stain shown in Fig. 4. Figure 5 shows the results, with Fig. 5a showing the absolute azimuthal CoM for all the benign glands for each patient. This value is then taken as a basis for all other (cancerous) gland types for each patient, thus providing a personalized reference point for a malignancy biomarker.

Figures 5b-c show the relative shifts in the CoM of cancerous glands relative to the benign glands of each patient (absolute shifts are shown in Fig. S3). A remarkable result of this personalized biomarker is that patients with the most aggressive form of prostate cancer (i.e., those containing Gleason grade 5 glands) exhibit a ubiquitous glandular phenotypical shift in the opposite direction as patients with less aggressive forms of prostate cancer. That is, for patients with aggressive cancer, both Gleason grade 5 regions and lower Gleason grade regions (3 and 4) show a negative azimuth CoM shift, which is the opposite behavior compared to patients with less aggressive tumors. This unique and ubiquitous shift—only present in aggressive prostate cancer—may be attributed to higher contributions from scattering which is indicative of changes in tissue organization at the nanoscale level. Similar behavior has been reported in other studies of the scattering properties of cancerous tissues (36-41) and is likely related to the field effect of carcinogenesis.

**UTOM for label-free H&E colorization with UV microscopy**

The label-free, quantitative UV images presented above can be used alongside the gold-standard H&E-stained tissue sections to improve diagnosis and grading, however it is also possible to translate the UV images into virtual H&E images. This has the advantage of providing standard, quantitative H&E images, free of cumbersome, time-consuming, and complex procedures, and avoids artifacts and variations that are common with H&E stains.

To translate the label-free UV images into virtual H&E images, we apply a recently developed *unsupervised content-preserving transformation for optical microscopy* (UTOM) deep neural network (57). UTOM adapts the general framework of cycle-consistent generative adversarial networks (Cycle-GAN) which can transform images from one domain into another without requiring pixel-level paired data. In UTOM, a forward and backward GAN are trained simultaneously to learn a pair of opposite mappings between the UV and H&E image domains, as shown in Fig 6a. In this process, a cycle-consistency loss constrain, and a pair of saliency constraints are imposed to correct for mapping direction, which avoids distortions (Fig. 6a) (57). In the training process, the overall network converges when the discriminators cannot differentiate between images produced by their generators (i.e., when the two GANs reach equilibrium; see Fig. 6b). Once trained, new images can be fed into the network and transformed into the desired domain (Fig 6c). This approach has been used for image restoration (e.g., resolution enhancement, removing distortions), for virtual fluorescence labeling of label-free phase images, and H&E virtual staining of autofluorescence images (58-60). The training set for this work comprised of 54 regions from 10 patients, while the test set (transformation group) contained 21 regions from the remaining 5 distinct patients. More details on the training process and final image translation are provided in methods and materials section.

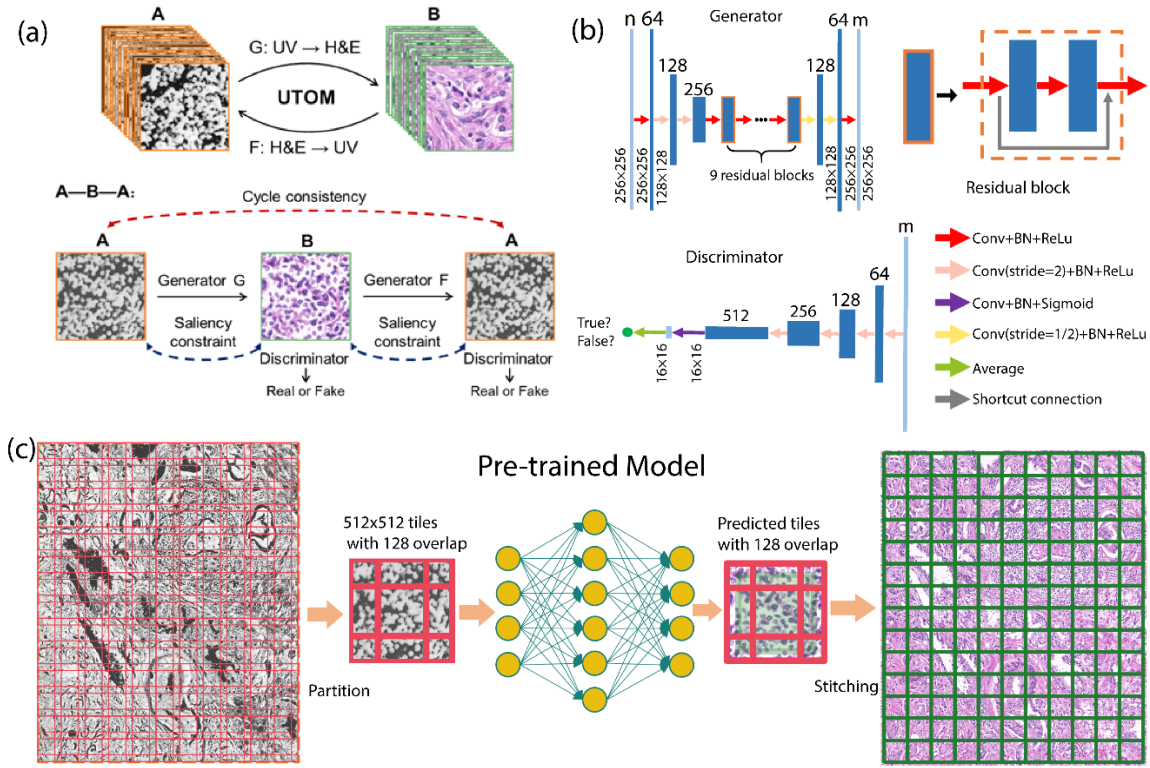

Figure 6: Schematic of colorization process and the UTOM method. For the transformation from UV to HE, input channels N=4, and output channels M=3. Each coral rectangle represents a feature map extracted by corresponding convolutional kernels. The generator is a multi-layer residual network with downsampling input layers and upsampling output layers. The discriminator (PatchGAN classifier) uses multiple strided convolution for abstract representation. It generates a matrix, in which each element corresponds to a patch in the input image. The ultimate output is the average of the loss over all patches

Figures 7a-d show two representative examples of UV translated (virtual H&E) prostate tissue images from the test set, along with their corresponding adjacent H&E-stained sections. The figures clearly show that the UV translated images are nearly identical to the H&E-stained tissues sections, with the most marked differences arising from the fact that the images are from adjacent sections. Specifically, the virtual H&E images preserve or improve several important features that play an important role in PCa diagnosis: First, as shown in Fig. 7b, the UV translated images successfully recapitulate the appearance of basal cells and basal cell lamina around benign glands which are of utmost importance for PCa diagnosis. This feature is also observed in Fig. 7l where an entrapped benign gland is clearly differentiated from surrounding cancer regions. Second, PCa regions shown in Figs. 7g ,7h, 7p and 7q depict luminal epithelial nuclei with more consistent (and arguably improved) contrast in the UV translated images compared to their corresponding H&E-stained sections. These types of structures are especially important in differentiating cancer glands from other mimickers of cancer where the structure of the gland is slightly disrupted. Finally, the appearance of the clear or pale eosinophilic cytoplasm as well as hyperchromatic nuclei are well preserved, which in some cases can be indicative of PCa Gleason Grade 4.

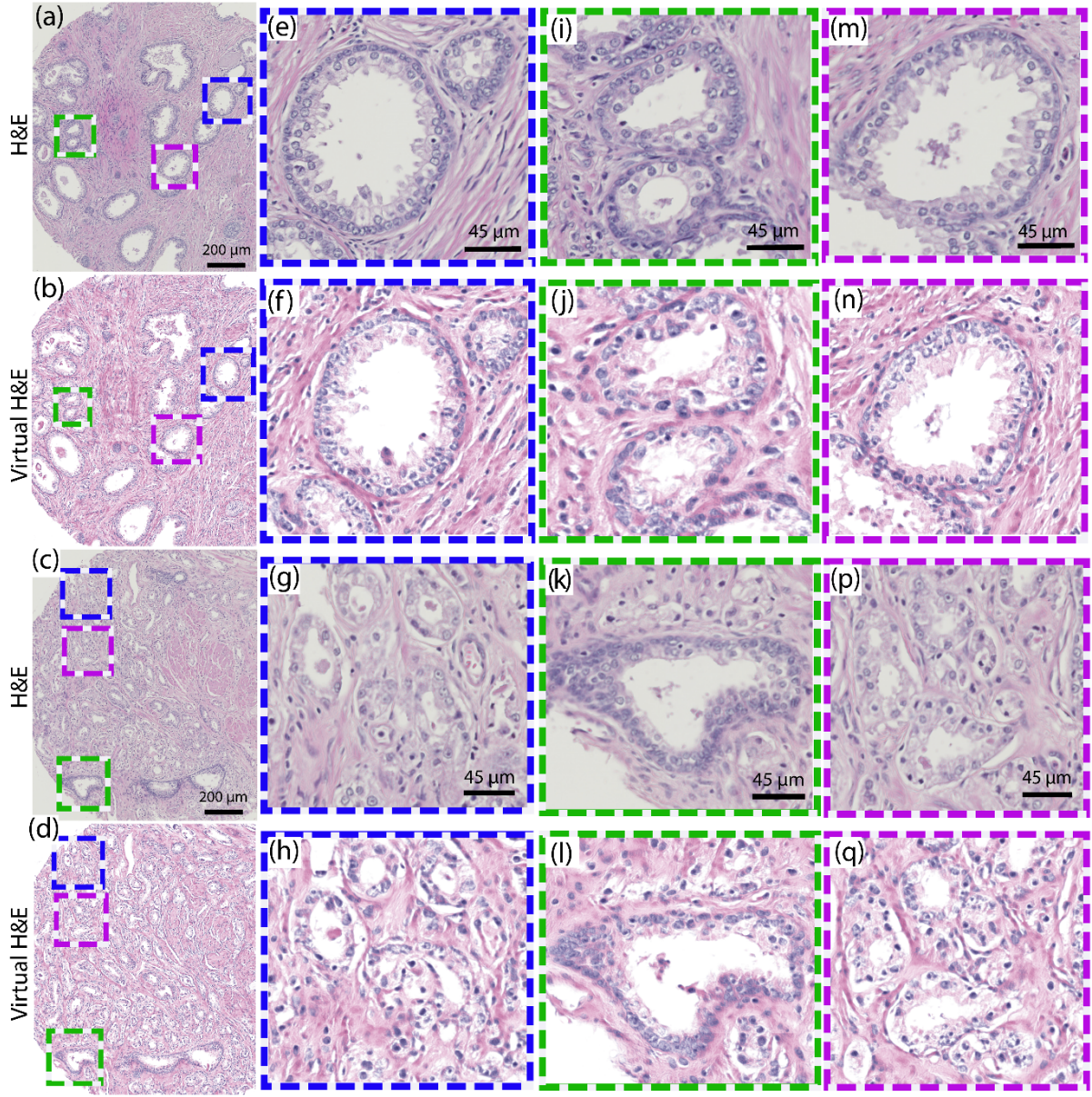

Figure 7: Examples of (b)-(d) Two predicted output virtual images (a) and (c) along with their reference H&E images. (e)-(q) show three selected zoomed regions for each area. These regions have been selected to compare features on both H&E and translated virtual H&E images.

**Virtual H&E evaluation**

To assess the quality of the UV translated, virtual H&E images compared to the gold standard H&E-stained images, we conducted a panel study with 4 board-certified/board-eligible histopathologists. Here the pathologists evaluated a total of 42 large area images (~1mmX1.5mm), half of the images (21) were images of H&E-stained tissue sections and the other half (21) were virtual H&E from the same regions (adjacent slices), all from the UTOM test set. Two pathologists (group 1) were assigned a set of 21 images comprising a mixture of virtual and stained H&E images, and the other two pathologists (group 2) were assigned the complimentary set, meaning images from the same regions but switching virtual H&E images with the images of stained H&E sections, and vice versa (no pathologist viewed the same region in virtual H&E and H&E-stained formats). While reviewing the images, pathologists were asked a series of questions regarding the quality of the images, with numerical scores ranging from 1 (poor) to 3 (excellent). They were also asked to provide a Gleason score for each region.

Results of the panel study are summarized in Table 1. Data show that the UV translated virtual H&E images and the H&E-stained tissue section images have very similar quality as assessed by the pathologist panel. With the exception of the nucleolus quality, which was evaluated slightly lower in the virtual H&E format, all other structures were assessed to have the same quality between the two modalities, with no statistically significant differences. The gland quality, which is of particular importance for PCa diagnosis, was deemed nearly identical between the two methods, as was the cytoplasm quality. Most importantly, the pathologists' diagnostic confidence was very similar for both methods (and not statistically different). We attribute the small difference in nucleus quality to the presence of lipid-laden macrophages (Xanthoma), mesonephric remnants, and hyperchromatic nuclei in a few regions with inflammation, which have a slightly disrupted visible quality in the translated images. However, (1) these are not diagnostically meaningful (which is likely why the diagnostic confidence remained the same between the two groups, even though the nucleus quality was slightly lower in the UV translated images), and (2) the nucleus quality can be improved with additional training.

**Table 1.** Comparison of translated UV images with H&E scans

| Parameter | H&E | Virtual H&E | Statistical significance |
|---|---|---|---|
| Nucleus Quality (scale of 3) | 2.57 | 2.32 | * |
| Cytoplasm Quality (scale of 3) | 2.45 | 2.42 | N.S. |
| Gland Quality (scale of 3) | 2.60 | 2.57 | N.S. |
| Diagnosis Confidence (scale of 3) | 2.26 | 2.16 | N.S. |
| Inter-group concordance for group 1 | 81.82% | 85% | - |
| Inter-group concordance for group 2 | 60% | 50% | - |
| N.S.: not statistically significant, * p-val <0.05 | | | |

We also calculate the inter-group concordance for grade group decisions using the H&E and virtual H&E images. The results (Table 1) show inter-observer variability at similar levels to what has been reported in previous studies (61-64). Importantly, however, the concordance in Gleason grade decisions is very similar between the UV translated virtual H&E images and the H&E-stained tissue section images *within each group*. These results strongly suggest that the format of the images (virtual H&E and H&E stained) did not play a role in the concordance levels. It is also worth noting that the agreement between the two most senior, board-certified pathologists was very high—they agreed in 17 out 21 regions even though they were viewing each region in different formats (one was in group 1 and the other in group 2). The 4 regions of discordance were between boarder line Gleason grades 3 and 4.

Finally, we calculate the accuracy of the Gleason scores provided by the pathologists for both the H&E and virtual H&E images. For this task, we first select a "ground truth" given by the decision of one of the two senior board-certified pathologists (whichever one of the two whose decision was based on the stained H&E section for a particular region was selected). Results show that the accuracy of the Gleason grades using H&E and virtual H&E images are similar (i.e., not statistically significant), with 72.5% accuracy for H&E and a slightly higher 77.45% accuracy for virtual H&E (p-value = 0.24). Alternatively, only using the regions where both senior, board-certified pathologists agree (17 out 21) and using their assessment as "ground truth", we find an accuracy for Gleason grade of 73.6% for H&E and 81.6% for the UV translated, virtual H&E images (again, differences not statistically significant, p-value=0.42).

**Discussion**

In this study, we have introduced multi-spectral deep UV microscopy as a novel, fast and reliable method to capture quantitative molecular and nano-scale information from unlabeled prostate tissue sections. We have utilized the unique UV spectral signature combined with an unsupervised spectral analysis to transform our multi-spectral data cubes into phenotypical maps or "optical stains" with subcellular spatial resolution. The spectral analysis suggests that the main contributing factors to these maps arise from scattering which serves as an indicator of tissue nano-architecture, and from proteins and nucleic acids. However, we do not rule out contributions from other molecules (35). Maps derived primarily from spectral signatures that correlate with proteins and nucleic acids provide high contrast among various critical tissue components, including nuclei, cytoplasm, basal layer, stroma, and glandular tissue, which can enhance our ability to recognize anomalies in prostate tissues.

While the "optical stains" derived from proteins and nucleic acids correlate well with the overall structures observed with the gold standard H&E stains, completely new structures are observed when incorporating the scattering signatures in conjunction with proteins and nucleic acids. These maps are likely indicative of micro and/or nano-scale alterations in the intracellular milieu, such as the cytoskeleton, ribosomes, chromatin, mitochondria, and collagen fibrils (36-41). Along with protein and nucleic acid alterations (43-51), changes in these structures have been implicated in the field effect of carcinogenesis. Indeed, here we observe that these structures map benign glands to different hues compared to cancerous glands, effectively yielding a "malignancy map." By quantifying these relative phenotypical shifts, we also find that cancer patients with the most aggressive forms of prostate cancer (those with Gleason grade 5 glands) possess a ubiquitous and unique phenotypical shift compared to patients with less aggrieve cancers.

These results have significant implications. Because less aggressive cancer glands (e.g., Gleason grade 3) possess a different phenotypical shift in patient harboring an aggressive cancer (those with Gleason grade 5 glands), this phenotype or biomarker may help identify patients with aggressive forms of prostate cancer even if initial biopsies miss the more aggressive regions. Together, all the "optical stains" may also help differentiate between border line cases as they provide additional molecular and biophysical information. For instance, quantitative information from the UV spectra and derived optical maps can help differentiate anomalous benign glands that mimic cancer and can be difficult to detect. Furthermore, all the images, supported by their histograms of the molecular signatures, show that healthy tissue, disease regions and their underlying composition span a continuum rather than a discrete distribution. This is in line with our understanding of disease progression(65-68) and may help better characterize prostate cancer compared to discrete labels (as with Gleason grades). This new information may also help assess a more ideal personalized treatment course for patients.

These results also lead to more fundamental questions: Do patients need to have this unique malignant phenotype to develop the aggressive form of PCa? If so, can it be detected even before aggressive cancer develops? How early? Or is there a ubiquitous switch across the gland once the disease progresses to this more aggressive form? The answer to these questions requires further understanding of this malignant phenotype and will guide our future work.

Finally, we have demonstrated the capabilities of our novel approach in digital pathology using a state-of-the-art deep learning-based virtual staining algorithm, UTOM. We showed that the UV images can be readily translated into virtual H&E images that accurately mimic the structures and colors present in the gold standard bright-field microscopy images of H&E-stained prostate tissue sections, without the need for laborious, time-consuming, and costly chemical staining procedures. Also, our UV translated images do not demonstrate staining viability and provide a consistent level of detail, colors, and contrast. A panel of board-certified/board-eligible pathologists assessed the quality and diagnostic potential of the UV translated images to be equivalent to the gold-standard H&E-stained tissue section images.

In conclusion, we have introduced multi-spectral deep UV microscopy as a novel quantitative tool for prostate tissue histopathology, which can help improve diagnosis and grading of prostate cancer. This tool provides novel insights based on the endogenous composition of tissues, while also providing the same critical information contained within today's gold standard (H&E) without the need for staining. The method is fast, low-cost, simple, and provides subcellular resolution, which overcomes critical limitations of other label-free optical methods applied to histopathology (15, 69-72). Finally, this same quantitative approach can be applied broadly across histopathological analysis of many tissue types and diseases.

**Materials and Methods**

**Deep-UV multispectral microscopy set up**

The deep UV transmission images were obtained using a microscopy system that consists of a plasma-driven broadband light source (Energetiq, EQ-99X) that provides a continuous spectrum from 200 nm to 2 μm. The output light from the source is focused on the sample using an off-axis parabolic mirror (Newport). A long-pass dichroic mirror is used to filter out the wavelengths of light above ~ 550 nm. For each region of interest, a multispectral data cube is captured using bandpass filters (bandwidth = 10nm) centered at 220, 255, 280 and 300 nm. The filters are placed on a filter wheel to change the imaging wavelength of the system. A 0.5 N.A. UV objective (Thorlabs LMU-40X-UVB) is used to collect the transmitted light and a biconvex (f=150 mm) lens is used to relay light onto a UV camera (PCO. Ultraviolet). A schematic of the setup is shown in Fig. S1. For each acquisition, the camera integration time was ~100 ms. Each captured region of interest represents a field of view of about ~170 μm × 230 μm. The resolution of our system is ~250 nm. In this work, we studied regions that were comprised of 64 tiles in the form of an 8 by 8 mosaic image. To enable reliable stitching, each tile has ~15% overlap with its neighbors. The final resulting region is approximately ~1 mm × 1.5 mm.

**Sample collection and preparation**

Paraffin-embedded formalin-fixed blocks from radical prostatectomy specimens were obtained from 15 prostate cancer patients. All the patients had not received any neoadjuvant therapy prior to radical prostatectomy. The Gleason scores (Grade groups) and tumor stages were assigned by

Urologic Pathologists in all cases. Next thin slices (~5 microns thick) of the tissue blocks were mounted on quartz slides and were deparaffinized by incubating the slides in Xylene bath for 5 minutes. The samples were then placed in 95% Ethanol for 3 minutes to remove Xylene and washed with dionized water. One section was used for UV imaging and a second section was stained with H&E and imaged with a bright field microscope.

All tissues are de-identified from archived tissue block for Emory University Hospital (n = 10) or a commercial vendor (Biomax) (n = 5). This work is conducted under an IRB except protocol (H16343).

**Data Processing**

To study the molecular content of the imaged tissue slides, different wavelengths in each captured multispectral data cube were registered in MATLAB (Mathworks) Environment. Next, in order to have a single wide-field UV image we used an image stitching code (MIST)(73), developed by National Institute of Standards to stitch the 64 tiles captured separately.

To calculate the principal components (PCs) of the multispectral prostate tissue images, we selected 90 regions that yielded approximately ~130 million spectra which represented all biologically important structures in prostate tissue. Next, we performed PCA in MATLAB to calculate the 4 principal components of the selected regions.

To generate color-coded images, we calculated the projections of the multispectral UV data on PC 1, 2, 3, and 4, respectively. Next, we converted the resulting projection vectors (Proj 1, Proj 2, Proj 3) and (Proj 2, Proj 3, Proj 4) from Cartesian coordinates to Spherical coordinates (Azimuth ($\theta$), Elevation ($\phi$), Radius (R)), where Proj i represents the projection of UV data on PCi. Finally, to get the geometrical representation of the PCA, we calculated a two-dimensional histogram of the azimuth ($\theta$) and elevation ($\phi$) angles for each case. Lastly, colorized the images using a Hue-Saturation-Value (HSV) color space, where the hues are assigned based on either azimuth or elevation angle, the value is set by the radius and the saturation is set to 1.

**Calculation of the azimuthal shifts**

To calculate the azimuthal shifts that are correlated with prostate cancer grades, first we annotated all the corresponding H&E images with appropriate Gleason grades. The annotations were reviewed and approved by a board-certified Urologic pathologist. Next, for each patient, the multispectral UV data were manually segmented according to the approved H&E annotations to extract all the pixel spectra that have the same Gleason grade. Once all the grade specified spectra were collected, we calculated cumulative 2D histograms using Principal components 1, 2, and 3 for each Gleason grade category as described in the data processing section. Finally, we integrated each 2D histogram in elevation direction to generate the azimuth dependent graph of molecular content, and recorded the Azimuth coordinates center of mass. We repeated this procedure for all the captured regions from all the patients.

**Virtual H&E Colorization using UV microscopy images**

To perform machine learning process, we used the label-free UV images of the unstained tissue sections from 15 patients from all 4 wavelengths (220,255,280 and 300 nm). For each captured region the corresponding H&E-stained image from adjacent slice were used as a reference. All the UV and H&E images were scaled to the same pixel size (90 nm). Next, we used 54 regions from 10 patients that contained representative biologically structures in prostate tissue, as the training data-set for our model (~13.5 billion spectra). The remaining regions (21) from the other 5 patients were used as the testing data set to evaluate the color transformation model. The important point about the testing data set is that the regions come from completely independent patients and no regions from testing patients are used in the training process. In the training dataset, the 4-channel UV data and H&E images (RGB channels) were randomly cropped into 512×512 patches. The total numbers of UV and H&E patches are 64336 and 81667, respectively (Fig. 6. (a)). During the test phase, the UV images were first partitioned into small patches with 25% overlaps. After a model was trained, patches from the previously unseen 5 patients were then fed into the model to generate the corresponding H&E patches. To finally form a large area virtual H&E image (each ~1mmX1.5mm), we cut out the boundaries (half of the overlap) of the generated patches and stitched the remaining parts together one by one.

**Virtual H&E Color normalization**

To remove undesirable color variations of the H&E-stained histological images, which result from differences in staining protocols, slide scanners and other factors, we adopted the structure-preserving color normalization (SPCN) method proposed by Vahadane et al (74). For a given image, we first estimated its stain density maps and color appearances via sparse non-negative matrix factorization. Then, we combined the stain density maps with a stain color basis of an arbitrary target H&E image so as to change only the color appearances while preserving the structure of the source image.

**UTOM method**

To produce virtual H&E colorized images, a forward GAN and a backward GAN are trained simultaneously to learn a pair of opposite mappings between two image domains. Along with the cycle-consistency loss, a saliency constraint is imposed to correct the mapping direction and avoid distortions of the image content. For each domain, a discriminator is trained to judge whether an image is generated by the generator or from the target domain (Fig. 6 (b)). When the loss converges, the two GANs reach their equilibriums, which means that the discriminators cannot distinguish images produced by their generators from the target images. An image could be mapped back to itself through the sequential processing of the two generators, and more importantly for biomedical images, the saliency map keeps high fidelity after each transformation (Fig. 6 (a)). The well-trained generator G of the forward GAN is used for transformation task from UV images to H&E images (Fig. 6(c)).

The architectures of the generator and the discriminator are visualized in Fig. 6b. The first three layers of the generator are downsampling layers implemented by strided convolution to extract low-level abstract representations. Nine stacked residual blocks are followed to extract high-level features. The number of residual blocks reflects the model capacity. More residual blocks are recommended for more complex tasks. The last three upsampling layers are also implemented by strided convolution. They are used to integrate extracted features and rescale the image to its original size. The discriminator is a relatively shallow CNN. Each layer downsamples the feature maps but doubles the channel number. The last convolution layer generates a single-channel feature map and classification is performed on each element of this feature map (PatchGAN classifier). The final true or false label is generated by averaging individual labels of all elements. Each convolution layer in both the generator and the discriminator contains a nonlinear activation unit. Whether to use the sigmoid function or rectified linear unit (ReLU) is marked with corresponding arrows in Fig. 6b.

The Adam optimizer was used to optimize network parameters (57). The initial learning rate is 0.0002, which decays linearly every 50 iterations with a rate of 0.99. The batch size was set to 1 and the images were flipped randomly for data augmentation. We trained the network for about 5 epochs, with about 80000 iterations in each epoch. On a single NVIDIA GEFORCE RTX 2080 Ti GPU (11GB memory), the whole training prcess took approximately 48h. After training, UTOM took 21ms to generate a 512x512 H&E patch and cost 3s to produce a whole-slide HE image.

We used a PC system with an Ubuntu 16.04 LTS operating system and a CPU Intel(R) Xeon(R) CPU E5-2683 processing unit. Also a PyTorch 1.6 was used as the Deep Learning Framework and Python 3.7 for image processing.

**Virtual H&E evaluation methodology**

We prepared a web-based survey including 21 unidentified, mixed H&E and virtual H&E regions (group 1, 10 H&E and 11 Virtual H&E and group 2, 11 H&E and 10 virtual H&E images of the same regions) and asked 2 board-certified and 2 board-eligible pathologists to submit their evaluations of the quality of parameters such as nucleus, cytoplasm and gland quality. Further, we asked them to submit a Gleason Score for each region to compare the accuracy of diagnosis for both H&E and virtual H&E images. Each question was based on the scale of 1 to 3 (1 for poor, 2 for moderate and 3 for very good quality). The responses were downloaded and used for statistical analysis. This clinical panel review protocol (no. H19389) was Institutional Review Board-exempt.

**Supplementary Materials**

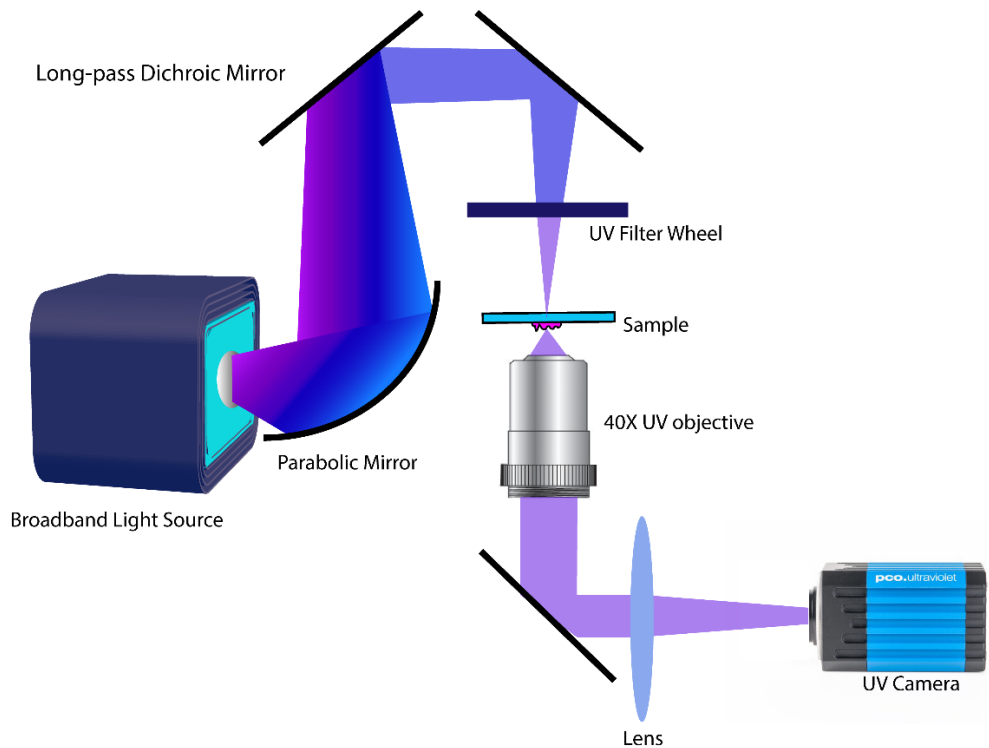

Fig. S1. A schematic of the multi-spectral deep UV microscope. The source provides a broadband output beam (~200 nm to 2000 nm) that is focused on the sample using a parabolic mirror. A dichroic mirror is used to only select deep UV region of the spectrum (200-550 nm). The transmitted light is collected using a 40X UV objective and is relayed on the camera using a biconvex lens.

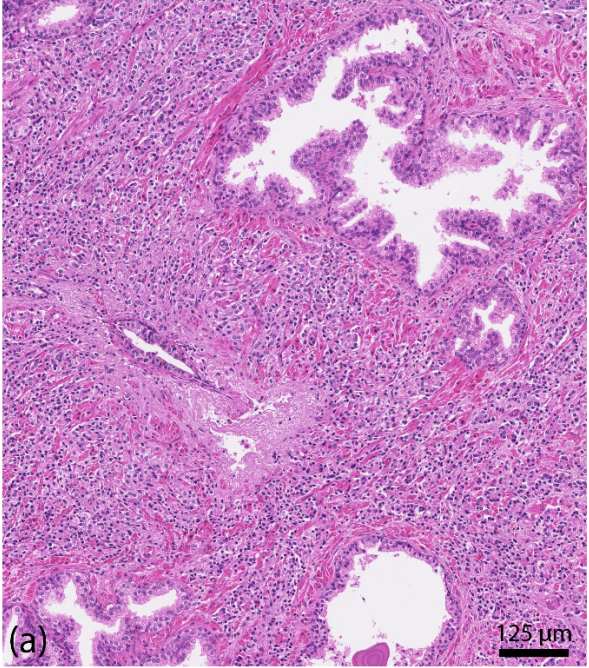
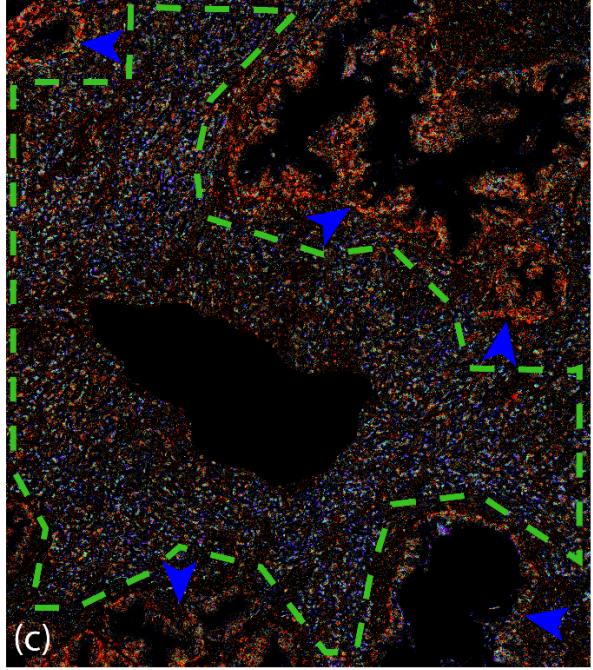
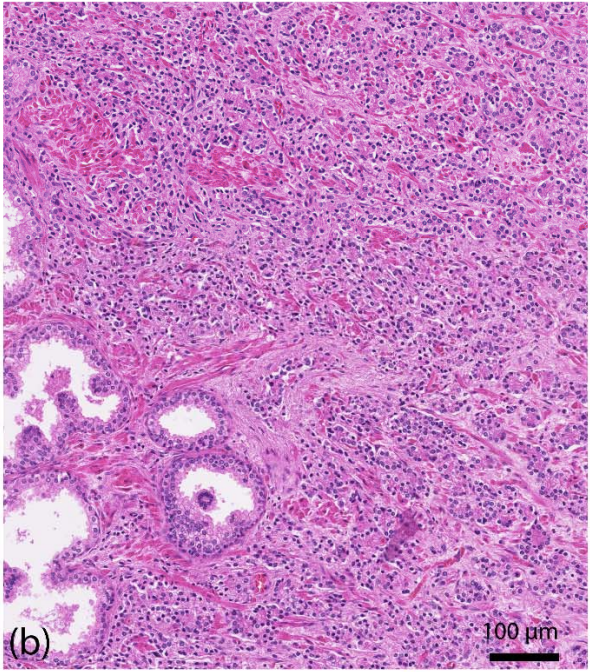
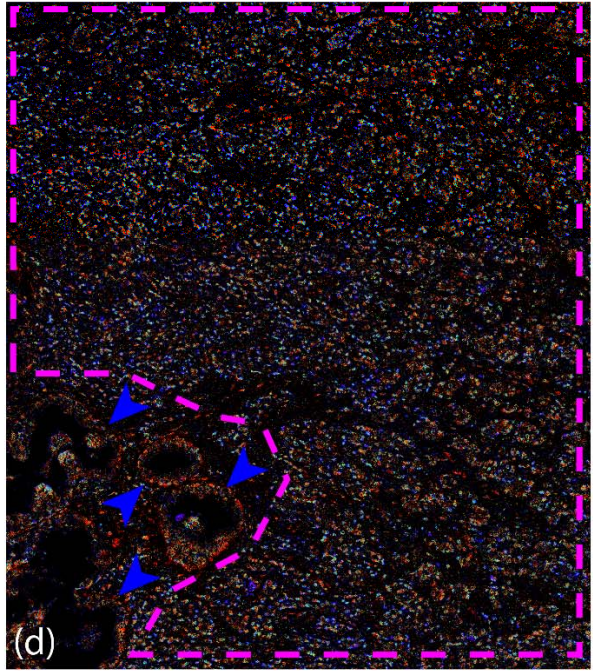

Fig. S2. Two examples of prostate cancer regions with aggressive forms of cancer (Gleason Grade 5). It is important to mention that the region in (b) starts from foci of Gleason Grade 4 in the right side of the image and gradually cancer regions with Gleason Grade 5 appear in the top left side of the region (there are not pure Gleason Grade 4 or Gleason Grade 5 regions). Some levels of inflammation are also present.

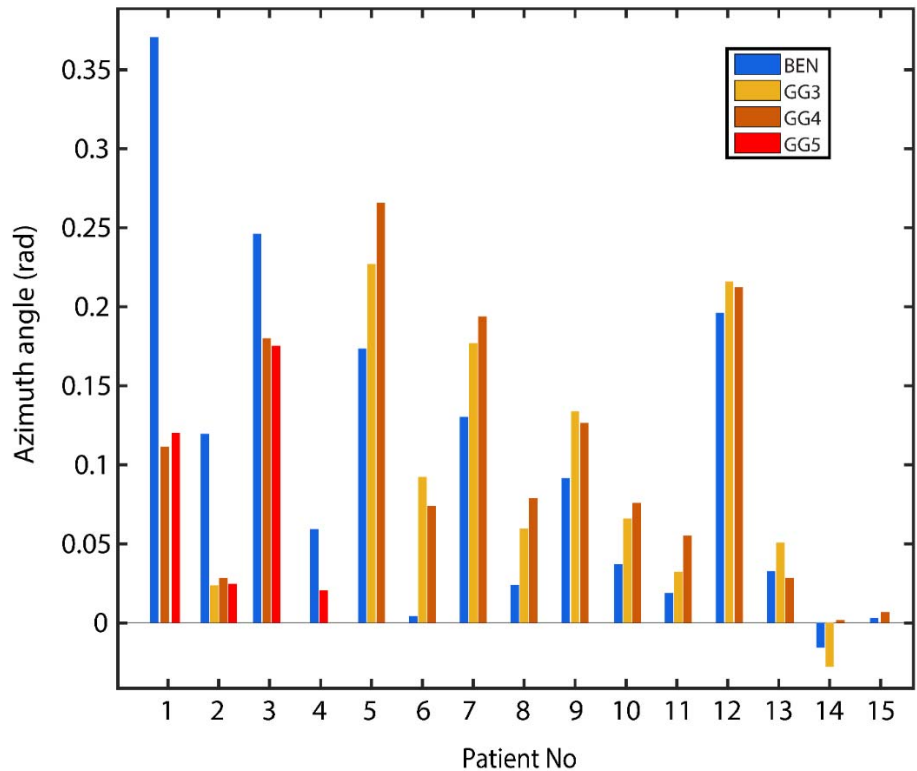

Fig. S3. Bar plot of the absolute azimuth angle CoM.

**Acknowledgments:** We thank Dr Fengming Chen, MD, PhD , Dr Patrick Mullane, MD and Dr Oluwaseun Ogunbona ,MD from Emory Hospital for evaluating the H&E translated UV images. We thank Steven Marzec for helping with the panel study web-page.

**Funding**: We greatly acknowledge support for this work from the Burroughs Wellcome Fund (CASI BWF 1014540); National Science Foundation (NSF CBET CA- REER 1752011); and the Wallace H. Coulter Biomedical Engineering Department at Emory University and the Georgia Institute of Technology.

**Author contributions**: S.S contributed to the experimental design, performed experiments and data acquisition, performed data processing and computational data analysis, histopathological analysis and evaluation and tissue annotation, wrote the manuscript and prepared the figures. A.O contributed to the experimental design and data acquisition. H.Q. developed the machine learning model and performed deep learning image translations. N.K. contributed in data acquisition. X.L. contributed to machine learning modeling. A.O.O. was the histopathology supervisor, approved prostate tissue annotations and contributed to translated UV image evaluation. Q.D. supervised the machine learning processes. F.E.R was project supervisor (Principal investigator), designed experiments, wrote the manuscript, and supervised all steps. All the authors contributed to final approval of the manuscript. F.E.R and S.S performed manuscript submission and review process.

**Competing interests** The authors declare that they have no competing interests.

**Data and materials availability**: The color-coded images are uploaded in: